# Magnetically-triggered Nanocomposite Membranes: a Versatile Platform for Triggered Drug Release

Todd Hoare[1†], Brian P. Timko[2,3†], Jesus Santamaria[4,5], Gerardo F. Goya[5], Silvia Irusta[4,5], Debora Lin[3], Samantha Lau[3], Robert Langer[3], and Daniel S. Kohane[2]*

[1]Department of Chemical Engineering, McMaster University, 1280 Main St. W, Hamilton, Ontario, Canada  L8S 4L7

[2]Laboratory  for Biomaterials and Drug Delivery, Department of Anaesthesiology, Division of Critical Care Medicine, Children's Hospital Boston, Harvard Medical School, 300 Longwood Ave., Boston, MA , U.S.A. 02115

[3]Department of Chemical Engineering, Massachusetts Institute of Technology, 45 Carleton St., Cambridge, MA, U.S.A. 02142

[4]Networking Biomedical Research Center of Bioengineering, Biomaterials and Nanomedicine (CIBER-BBN). Maria de Luna, 11. Zaragoza 50018 Spain

[5] Institute of Nanoscience of Aragón, University of Zaragoza, Pedro Cerbuna 12, 50009 Zaragoza, Spain.

[†]These authors contributed equally to this report.

*To whom correspondence should be addressed. E-mail:
Daniel.Kohane@childrens.harvard.edu

**Abstract**

Drug delivery devices based on nanocomposite membranes containing thermoresponsive nanogels and superparamagnetic nanoparticles have been demonstrated to provide reversible, on-off drug release upon application (and removal) of an oscillating magnetic field. The dose of drug delivered can be tuned by engineering the phase transition temperature of the nanogel, the loading of nanogels in the membrane, and the membrane thickness, allowing for the delivery of drugs over several orders of magnitude of release rates. The zero-order kinetics of drug release through the membranes permit drug doses from a specific device to be tuned according to the duration of the magnetic field. Drugs over a broad range of molecular weights (500-40,000 Da) can be delivered by the same membrane device. Membrane-to-membrane and cycle-to-cycle reproducibility is demonstrated, suggesting the general utility of these membranes for drug delivery.

Sustained drug release technology has been applied in a wide variety of medical fields[1]. Many devices are passive, exhibiting release kinetics that are either constant or decreasing over time. However, drug delivery devices that can be repeatedly switched on and off would be optimal for effective treatment of conditions such as diabetes, chronic pain, or cancer.[2] To this end, environmentally responsive ("smart") materials have been developed that can respond to stimuli that are either internal to the patient (e.g. body temperature) or external (e.g. a remotely-applied magnetic field). Temperature-sensitive drug delivery devices have been developed based on the thermoreversible polymer poly(N-isopropylacrylamide) (PNIPAm)[3], which has been incorporated into implantable hydrogels[4-9], microparticles[10] nanoparticles[11-14], and surface-grafted polymers[15-27]. Examples of magnetically-activated materials include superparamagnetic nanoparticles, which heat and/or vibrate when placed in an oscillating magnetic field and have been used to achieve drug release from polymer scaffolds[28], sheets[29], liposomes[30], microspheres[31, 32], microcapsules[33], and nanospheres[34-36], typically by mechanical disruption of the drug-biomaterial matrix. However, the quantity of drug contained by most of these "smart" carriers is relatively small, and drug release is characterized either by a single burst event or inconsistent dosing as a function of triggering cycle.

To achieve both triggered drug release and consistent dosing, we previously reported composite membranes containing both temperature-sensitive and magnetically activated components.[37] These membranes were used to contain reservoirs of drug and achieved repeatable, on demand, on-off switching of molecular flux upon application of an oscillating magnetic field. However, in order to successfully translate this technology to the clinic, factors affecting the basal (off-state) release rate, on/off ratio, and drug-membrane interactions need to be understood, and rationally controlled.

In the present study, we report, for the first time, the relation between the chemical and physical composition of the membranes and the release kinetics of a variety of model compounds. Specifically, we demonstrate (a) how the critical on-off temperature of the device can be controlled by the chemical composition of the polymer nanogel, (b) how drug release can be tuned as a function of nanogel loading density and membrane thickness, and (c) how these membranes can be used to deliver both small and large, and anionic and cationic molecules.

Membranes were produced by suspending or dissolving superparamagnetic iron oxide nanoparticles, PNIPAm-based nanogels (NGs), and ethyl cellulose (the membrane matrix material) in ethanol and evaporating to form a film (Figure 1a-c; see Supporting Information for Methods). We have hypothesized that the nanogel forms a disordered, interconnected network througout the matrix. The superparamagnetic nanoparticles behave as local heat sources that are activated (turned on) by an external, oscillating magnetic field.[38] Temperature-triggered collapse of the PNIPAm enables transport of material (i.e. drug molecules from an enclosed reservoir) across the membrane (Figure 1d).

We produced membranes containing various types and quantities of nanogels prepared to exhibit different phase transition temperatures. The phase transition temperature was controlled by copolymerizing NIPAm with N-isopropylmethacrylamide (NIPMAm) and acrylamide (AAm) (Table 1). Figure 2a indicates that copolymerization of different combinations of precursors can shift the transition temperature of the nanogels from ~32°C (NG32) to ~46°C (NG46). Of particular note, using this copolymerization approach, the transition temperature can be shifted without inducing a change in the total percentage volume change observed upon nanogel deswelling (Table 2, $p>0.2$ for all pair-wise comparisons). As a result, highly thermoresponsive nanogels can be synthesized which have a range of phase transition temperatures appropriate for a variety of different triggering applications.

The transition temperature of molecular flux through a membrane correlates well with the phase transititon temperature of its consitutent nanogels, as shown in Figure 2b. However these data also indicate a temperature offset between flux and transition temperature. For example, with the NG32 membrane at 30°C, the nanogel shrank by ~250 nm without the occurrence of a significant change in permeability. A similar trend is observed with the NG37 membrane, with a ~200 nm size decrease required prior to a significant increase in membrane permeability. This lag may be attributable to the presence of a disordered pore network inside the membranes; small volume changes in the nanogel cannot generate sufficient free volume over the full thickness of the membrane to significantly change the diffusion coefficient of drug through the membrane.

To change the flux of drug through the membrane in the on state, the diffusional resistance to drug flux across the membrane must be controlled and optimized. This can most directly be accomplished by changing the porosity of the membrane or changing the thickness of the membrane. Membrane thickness is the easiest method for adjusting drug flux. Increasing the membrane thickness increased the diffusional path length of a drug molecule through the membrane and thus reduced the rate of drug release (Figure 3). The thinnest membrane tested (90 ± 14 µm thick) released 6.4 ± 0.4 µg/hr sodium fluorescein while the thickest membrane tested (288 ± 32 µm thick) released only 0.4 ± 0.1 µg/hr sodium fluorescein. Mass transfer rate correlated well ($R^2$ = 0.92) with membrane thickness measured with callipers. Therefore, a trade-off existed between membrane strength and membrane flux; thicker membranes that are presumably stronger release drug more slowly. Drug release rate could be further tuned by adjusting the concentration gradient of drug across the membrane, as a linear correlation exists between the initial drug concentration and the membrane flux (Supporting Figure S1).

Increasing the nanogel content inside the membrane increased the number of thermoresponsive pores templated into the membrane and thus increased the total free volume generated inside the membrane when the nanogels underwent a phase transition, leading to significant increases in drug flux through the membrane (Figure 4). Indeed, at nanogel loadings of 25wt% or less, a logarithmic relationship existed between membrane flux and nanogel loading (Figure 4B, $R^2$ > 0.94). As a result, the mass flux of sodium fluorescein through the membrane could be tuned over at least two orders of magnitude (0.1-10 µg/hr) by changing the nanogel content inside the membrane. The same trend was observed when an oscillating magnetic field was used as the on-off trigger (Figure 5); the membrane containing 23 wt% nanogel exhibited a lower average drug release rate in the on state (4.1 µg/min) than did the 28 wt% nanogel membrane (5.7 µg/min). In this case, both membranes were heated by the same ~2.2°C temperature gradient under the applied oscillating magnetic field, due to the identical ferrofluid content in each membrane. The 23 wt% membrane also had a longer induction time to the on state (~35 minutes versus ~15 minutes for the 28 wt% nanogel membrane) and returned to the baseline (off) flux level more slowly (~15 minutes lag time versus < 1 minute for the 28 wt% nanogel membrane), suggesting that nanogel content affected the kinetics of the on-off transition in the composite membranes.

However, the ratio between the flux in the on state and the off state (i.e. the flux selectivity between the on and off states) decreased as the amount of nanogel in the membrane increases, particularly at higher nanogel loadings. The ratio between sodium fluorescein flux at 45 °C (on) and 37 °C (off) was 15.2 ± 2.6 for a membrane containing 12 wt% nanogel, 8.1 ± 1.5 for a membrane containing 25 wt% nanogel, and 6.0 ± 0.5 for a membrane containing 32 wt% nanogel. Thus, while increasing the nanogel concentration consistently increased the drug flux through the membrane, increased flux was accompanied by decreased on-off resolution. Indeed, membranes prepared with 37 wt% nanogel exhibited high flux in both the on and off states regardless of temperature, representing a "leaky" system that would be less useful for on-demand drug delivery (data not shown).

At all different membrane thicknesses and for all nanogels tested, drug release occurred with zero-order release kinetics over at least 24 hours ($R^2 > 0.98$ in all cases in Figures 3 and 4) in the on state. Therefore, the total dose of drug delivered over time could be dynamically adjusted by varying the duration of the oscillating magnetic field. Zero-order release was also observed in the off state, although the rate of drug release is significantly reduced (Supporting Figure S2). Thus, while the nanogel loading and/or membrane thickness could be engineered to control the magnitude of the drug release rate targeted for a particular membrane device, the duration of the on pulse could be used to precisely control the total amount of drug delivered using any specific device, providing full on-demand control (at both the design stage and in the patient use stage) over drug delivery.

On-demand, zero order release was also achieved for the flux of drugs with a range of physical properties. Triggered release was demonstrated from a saturated solution of a 40 kDa molecular weight fluorescein-labelled dextran (Figure 6). Effective on-off switching of macromolecule release was observed, with 0.28 ± 0.08 µg/hr drug flux measured through the membrane in the on state and a flux ratio of 6.7 ± 1.2 between the on and off states. In comparison, sodium fluorescein release through the same membrane occured at a rate of 8.0 ± 2.8 µg/hr at a much lower concentration gradient (1.25 mg/mL, see Methods in Supporting Information) with a flux ratio of 8.0 ± 1.5 between the on and off states (see Figure 4). Therefore, the lower permeability (in terms of absolute flux) of the membrane to FITC-dextran was likely due to its higher molecular weight. However, the similar flux ratio observed between the large and small molecules suggests that the flux ratio was predominantly governed by the inherent properties of the membrane (e.g. nanogel loading and thickness).

Thermally-triggered drug release was also demonstrated for bupivacaine, a small molecule amphiphilic drug that is largely cationic at physiological pH (Supporting Figure S3). Thus, it was possible to deliver drugs with different molecular weights and different charges using the same membrane-based delivery vehicles.

Successful use of these membranes as long-term, on-demand drug delivery vehicles also demands high reproducibility of cycle-to-cycle and device-to-device drug release. A representative example of four replicate runs for the 25 wt% NG37 membrane is shown in Supporting Figure S4. For a single membrane, the cycle-to-cycle variability is low; indeed, there is no statistical difference in drug release in either the on or off states on a cycle-to-cycle basis for any membranes tested in this work ($p > 0.07$ for any pair-wise comparison over four on-off cycles). Thus, a single membrane gives highly reproducible release profiles upon multiple triggering events. Similar results were observed over 10 triggering cycles when the membranes were fabricated into implantable reservoir drug delivery devices, although a slight lag in release was observed in the first on cycle, likely due to the need to first saturate the microgel-filled pores of the membrane with the drug prior to drug release. (Supporting Figure S5)

Membrane-to-membrane variability was low for the highly nanogel-loaded membranes ($p > 0.18$ for any pair-wise comparison between 32 wt|% nanogel-functionalized membranes, Figure 4A) but increased for membranes with lower nanogel loadings ($p < 0.05$ for at least one pair-wise comparison of on state releases for all membranes prepared with nanogel loadings of 25wt% or less, Figures 4A and S5). Observed membrane-to-membrane variability was likely attributable to subtle differences in hand-mixing of the highly viscous precursor solution/suspension between different membranes, leading to slightly different nanogel distributions in replicate membranes. Automation of the mixing procedure could minimize this variability.

The composite membrane-based drug delivery devices described here offer a significant improvement over existing technologies since they can be readily engineered to achieve rational control over drug release kinetics for a variety of compounds. Our data show that it is possible to precisely control drug dosing over multiple orders of magnitude of dosings based on both the physical properties and compositions of the membrane and the duration of the on pulse applied to a given membrane. The frequency and power of the applied magnetic field could also be tuned to effect changes in drug dosing by changing the steady-state temperature of the device. Furthermore, our devices exhibit excellent reproducibility device-to-device as well as cycle-to-cycle. This tunability and reproducibility would present multiple options to the engineer, clinician, and patient for dynamically changing the specific level of basal drug release as well as on-demand drug dosing.

**Acknowledgements:** This research was funded by NIH grant GM073626 to DSK. TH acknowledges post-doctoral funding from the Natural Sciences and Engineering Research Council of Canada. BPT acknowledges a Ruth L. Kirschstein NRSA fellowship, NIH Award Number F32GM096546. SI and GFG acknowledge support from the Spanish MEC through the Ramon y Cajal program.

**Figures**

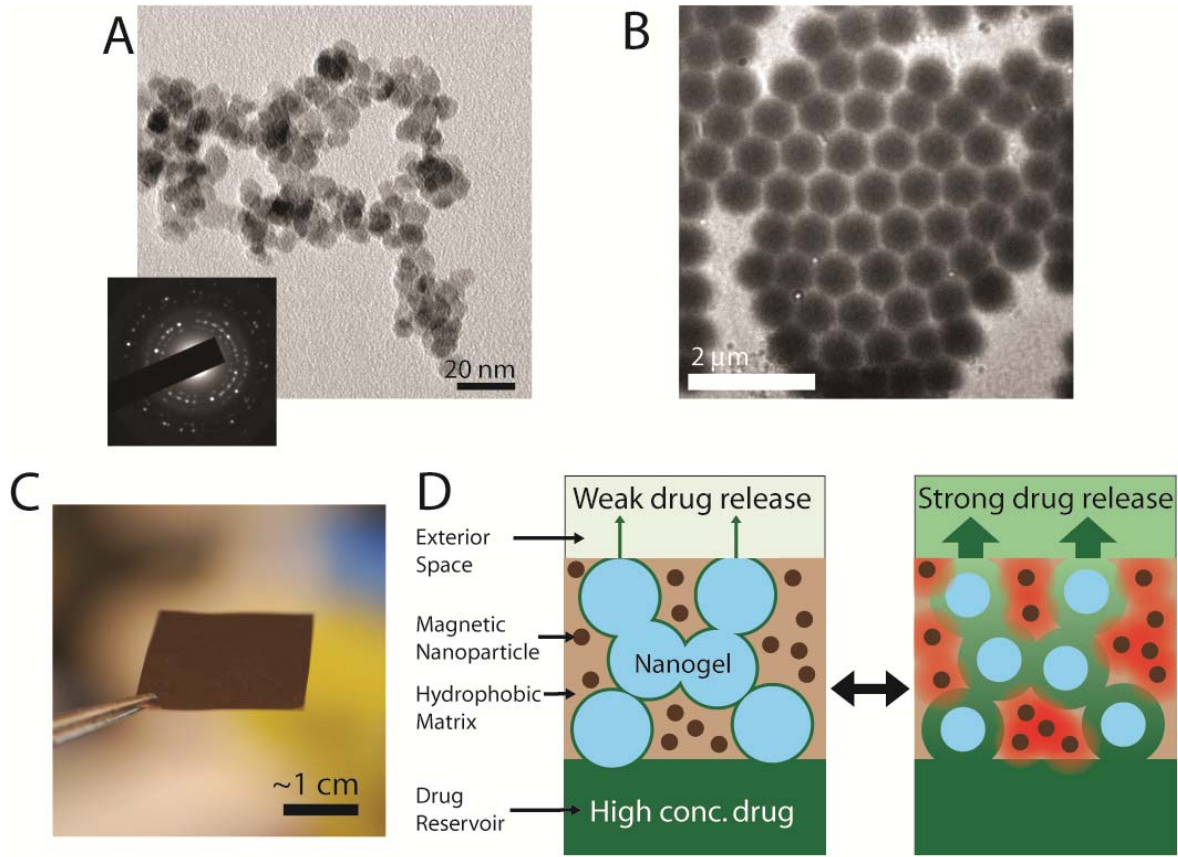

**Figure 1. Overview of membrane composition and function.** (A) TEM of superparamagnetic iron oxide nanoparticles. Inset: Diffraction pattern suggests an ensemble of randomly-oriented crystalline particles. (B) TEM image of dehydrated nanogel particles. (C) Photograph of membrane containing ferromagnetic nanoparticles and nanogel, distributed throughout an ethyclellulose matrix. (D) Proposed schematic of a cross-section of the membrane, showing nanogel particles (blue), iron oxide nanoparticles (dark brown), and ethylcellulose matrix (light brown). Upon application of a magnetic field, the magnetic nanoparticles release heat (red) and reversibly shrink the nanogel, enabling release of a drug (green) from a reservoir contained by the membrane.

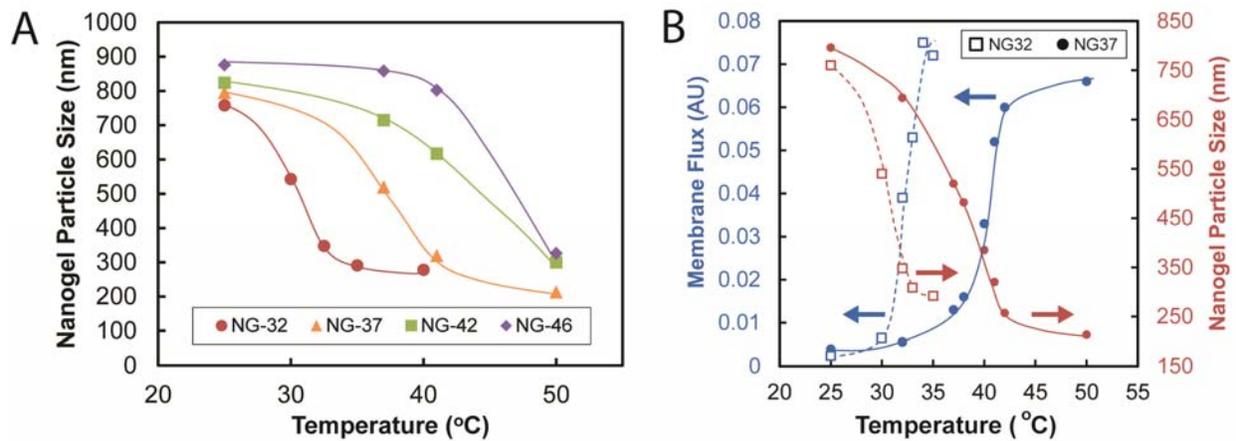

**Figure 2. Membrane on-off temperature can be tuned by nanogel composition** (A) Volume phase transition behavior of four nanogel formulations containing differing amounts of NIPAm, NIPMAm, and AAm (see Table 1). (B) Correlation between nanogel particle size in suspension (red points / right axis) and the mass flux (blue points / left axis) of sodium fluorescein through membranes as a function of temperature, for membranes containing 25 wt% NG32 and NG37 nanogels.

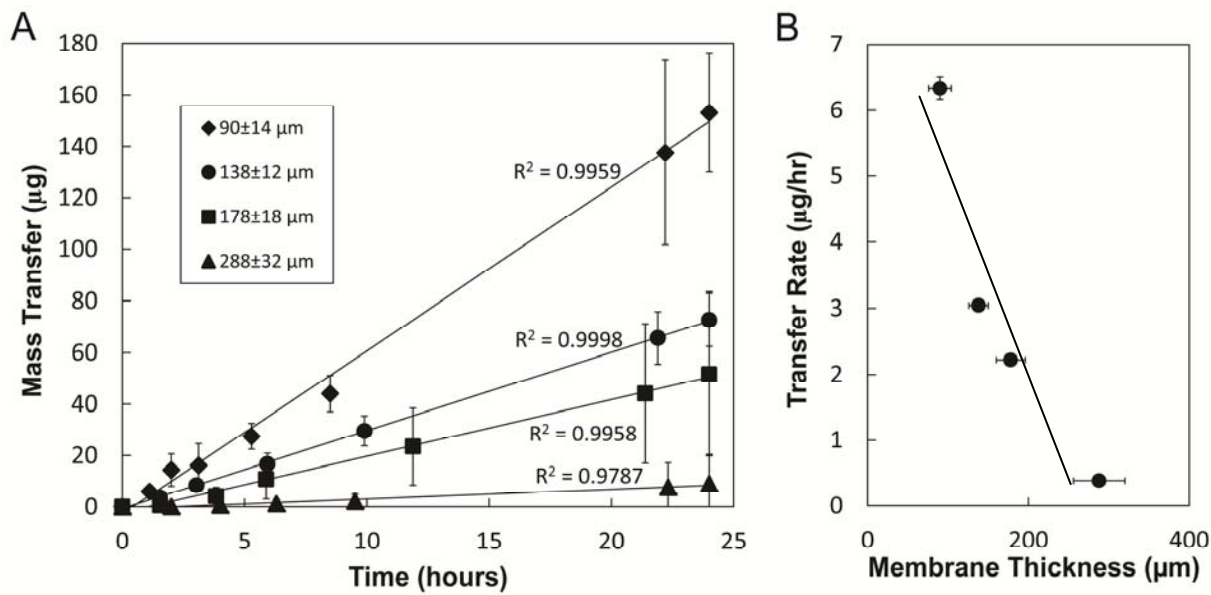

**Figure 3. Membrane thickness regulates sodium fluorescein flux.** (A) Mass transfer of sodium fluorescein as a function of "on" triggering time for membranes with different thicknesses. (B) Rate of mass transfer as a function of membrane thickness for data represented in panel A. All data are for sodium fluorescein flux, 25 wt% NG-37 membranes. Data are means ± SD; for each set, n = 6.

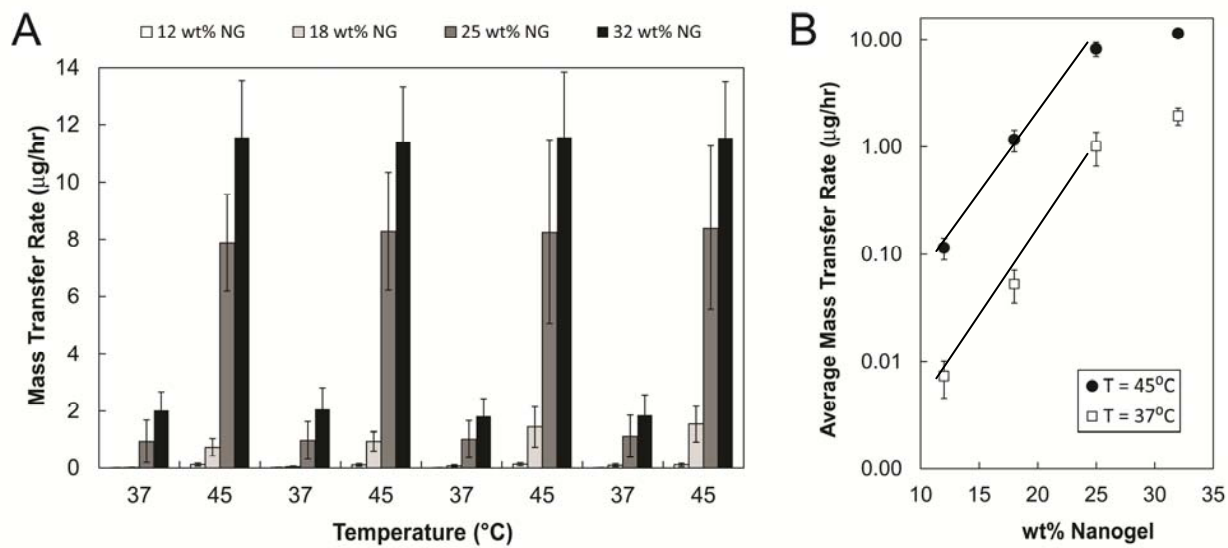

**Figure 4. Nanogel content of membranes regulates membrane flux**. (A) Repeated "on" and "off" cycles with temperature triggering at ca. 12-hour time intervals. (B) Average mass transfer rates of all cycles represented in panel A. Note the logarithmic scale. Data are for sodium fluorescein flux, NG-37 membranes. Data are means ± SDs; n =5, 4, 6, 6 for 12, 18, 25 and 32 wt% membranes, respectively.

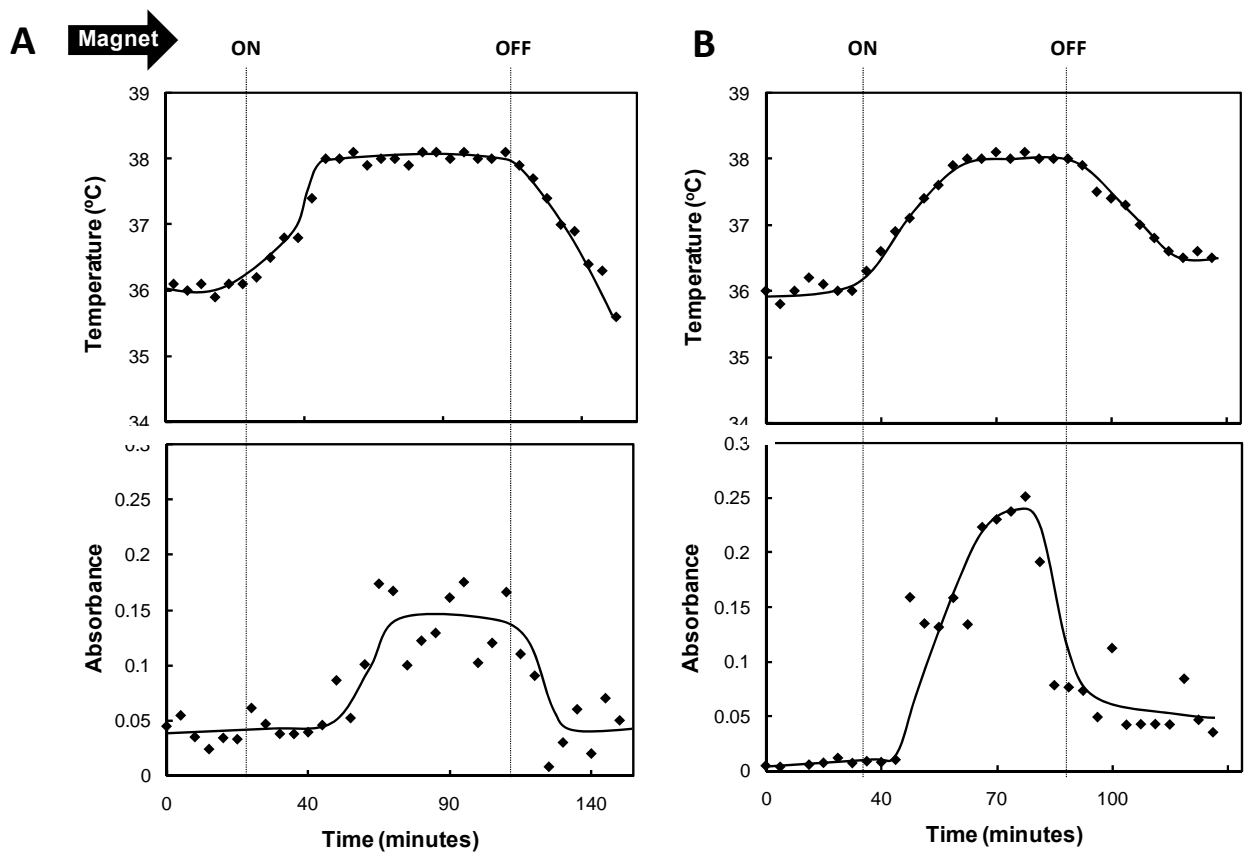

**Figure 5. Magnetic triggering of membranes.** Devices filled with sodium fluorescein and capped with membranes were turned "on" by an oscillating magnetic field (220-260 kHz, 0-20 mT). We separately measured devices containing (A) 23 wt% NG-37 or (B) 28 wt% NG-37 membranes.

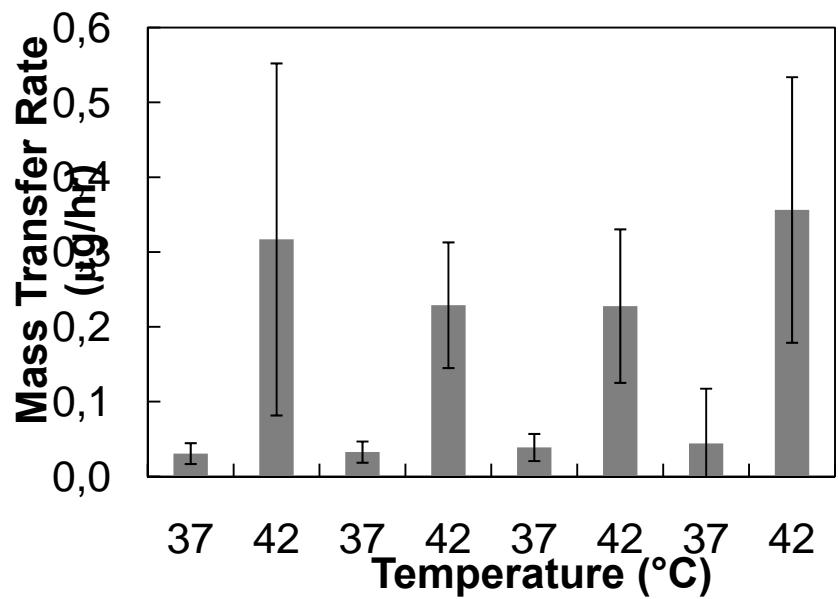

**Figure 6. Membranes can deliver large molecular weight molecules.** Mass rate of drug release through nanogel-filled magnetic membranes as a function of temperature (fluorescein-labelled dextran (40kDa molecular weight), 1.25mg/mL source solution, 25 wt% NG-37 membrane, thermal stimulus, ca. 4-hour time intervals. Data are means ± SDs; n=5.

**Tables**

Table 1.  Composition and thermal phase transition temperatures of four nanogels synthesized for membrane testing. Each membrane also contains 5 mol% N,N-methylenebisacrylamide crosslinker.

| Nanogel [a] | N-isopropylacrylamide (NIPAm, mol%) | N-isopropylmethacrylamide (NIPMAm, mol%) | Acrylamide (AAm, mol%) | Transition Temperature (°C) |
|---|---|---|---|---|
| NG-32 | 100 | 0 | 0 | 32 |
| NG-37 | 54 | 35 | 11 | 37 |
| NG-42 | 35 | 58 | 7 | 42 |
| NG-46 | 34 | 55 | 11 | 46 |

[a] The nanogel number refers to the transition temperature (right column).

Table 2.  Volume change (% volume change) on deswelling for the four nanogels synthesized for membrane testing

| Nanogel | % Volume Change on Deswelling |
|---|---|
| NG-32 | -95.1 ± 2.9 |
| NG-37 | -98.0 ± 2.5 |
| NG-42 | -95.2 ± 2.2 |
| NG-46 | -94.8 ± 2.8 |

*Supporting information for:*

**Engineering externally-triggered nanogel-magnetite composite membranes for drug delivery**

Todd Hoare, Brian P. Timko, Jesus Santamaria, Gerardo F. Goya, Silvia Irusta, Debora Lin, Samantha Lau, Robert Langer, and Daniel S. Kohane

**Methods**

**Supporting Figures S1 - S5**

**Methods**

*Materials:* N-isopropylacrylamide (NIPAm, 99%), N-isopropylmethacrylamide (NIPMAM, 97%), N,N-methylenebisacrylamide (MBA, 99%), acrylamide (AAm, 99%), ammonium persulfate (APS, 99%), iron (III) chloride (97%), iron (II) chloride (98%), ammonium hydroxide (28% in water), poly(ethylene oxide) (8000Da molecular weight), ethyl cellulose (97%), and ethanol (100%) were acquired from Sigma-Aldrich. Sodium fluorescein (99%), bupivacaine hydrochloride (99%), and FITC-dextran (molecular weight 40 kDa) were also acquired from Sigma-Aldrich. Water was of Milli-Q grade.

*Nanogel Synthesis:* Nanogels with different phase transition temperatures were prepared via copolymerization of N-isopropylacrylamide, N-isopropylmethacrylamide, and acrylamide using the recipes shown in Table 1. The numbers in the microgel code used represent the measured volume phase transition temperatures of the respective microgels. In each case, the monomers together with 0.08 g MBA (crosslinker) were dissolved in 150 mL water in a 500 mL round-bottom flask equipped with a magnetic stirrer, purged with nitrogen for 30 minutes, and heated to 70 °C under 200 RPM mixing. The reaction was initiated by adding a solution of 0.1 g ammonium persulfate in 5 mL of water. The reaction proceeded for at least 4 hours, after which the nanogel suspension was cooled, purified by dialysis using a 50 or 500 kDa MWCO membrane, and lyophilized. Nanogel diameters were measured using a ZetaPALS instrument (Brookhaven Instruments). Critical temperature is defined as the temperature at which the nanogel diameter is halfway between that of the fully swollen and fully collapsed state.

*Ferrite Nanoparticle Synthesis:* 3.04 g of $FeCl_3$ and 1.98 g of $FeCl_2$ were dissolved in 12.5 mL of distilled water. Ammonium hydroxide (6.5 mL) was added dropwise under 500 RPM mixing over 10 minutes. After 10 additional minutes of mixing, 1 g of PEO dissolved in 10 mL of water was added and the mixture was heated to 70 °C for 2 hours to peptize the ferrofluid surface. The ferrofluid was then cooled, washed using magnetic separation against distilled water (5 cycles), and concentrated to ~15 wt%.

*Microscopy:* Transmission Electron Microscopy (TEM) was performed on a JEOL 2100 TEM. Ferrite nanoparticles and nanogels were imaged at acceleration voltages of 200 kV and 80 kV, respectively.

*Membrane Formulation:* Most membranes were formulated by mixing 1.3 g of 10 wt% ethylcellulose in ethanol with the required amount of a 60 mg/mL nanogel suspension to achieve the targeted nanogel loading in a membrane. The ethylcellulose and nanogel components were mixed in a 60mm diameter Petri dish. The aqueous concentrated ferrofluid suspension was then pre-mixed with an equal volume of ethanol and added dropwise to the ethylcellulose-nanogel mixture to reach the desired magnetite concentration. Most membranes reported in this paper contained 20 wt% dry ferrite nanoparticles, with the exception of those reported in Figure S5, which contained 7 wt% dry ferrite nanoparticles. For thickness variation experiments, membranes were fabricated by mixing 0.9, 1.3, 1.8, or 2.3 g of the 10 wt% ethylcellulose solutions with nanogels and ferrofluid as required to achieve appropriate concentrations. In all cases, the membrane precursor suspension was mixed to homogeneity. Ethanol was evaporated over 2-3 days, or for as long as a week, to generate the dry membranes.

*Thermal Triggering:* Membrane flux was assayed by compressing a membrane between two glass flow cell chambers (side-bi-side cells, PermeGear, Inc.) filled with phosphate buffered saline and submerged in a water bath. The target chemical (1.25 mg/mL sodium fluorescein solution or 1.25 mg/mL FITC-dextran) was then added to one side of the flow cell. After pre-determined time intervals, samples were taken from the receiving chamber of the flow cell. For sodium fluorescein and FITC-dextran, the absorbance of the samples was then measured in a 96-well polystyrene plate using a multiwell plate reader operating at 490 nm. In all cases, the measured absorbances were converted to concentrations based on comparison to a calibration curve constructed from standard solutions.

*Magnetic Triggering:* Two 1cm diameter membrane disks were glued (using a Lock-Tite low viscosity adhesive) to the ends of a 1cm length of 3/8" OD (1/4" ID) silicone tubing containing a solution of 100 mg/mL fluorescein in saline. The drug-loaded membrane device was placed into a 2 mL Teflon reservoir inserted into a semi-adiabatic sample space in the centre of a water-cooled, seven-turn copper solenoid induction coil with a 4 cm internal diameter and 10 cm height. Continuous water flow through the sample at a rate of 2.2 mL/min enabled sampling of the fluorescein flux from the device as a function of time in the presence of an oscillating magnetic field of frequency 220-260 kHz and field amplitude 0-20 mT. Temperature was measured using a fiber optic probe located inside the sample holder. Samples were analyzed for sodium fluorescein concentration using an Agilent 8453 UV/VIS spectrophotometer operating at 487 nm.

*Error Analysis:* All error bars represent one standard deviation. For each data set that includes error analysis, six identical membranes were evaluated in parallel. A few membranes produced anomalous data, potentially because of occlusion of the membrane by an air bubble (abnormally low flux) or leakage of the membrane or seal (abnormally large flux). Such membranes were excluded from the data sets. On-off flux ratio errors are reported as standard errors of the mean (4 cycles per membrane, 6 membranes tested per membrane composition).

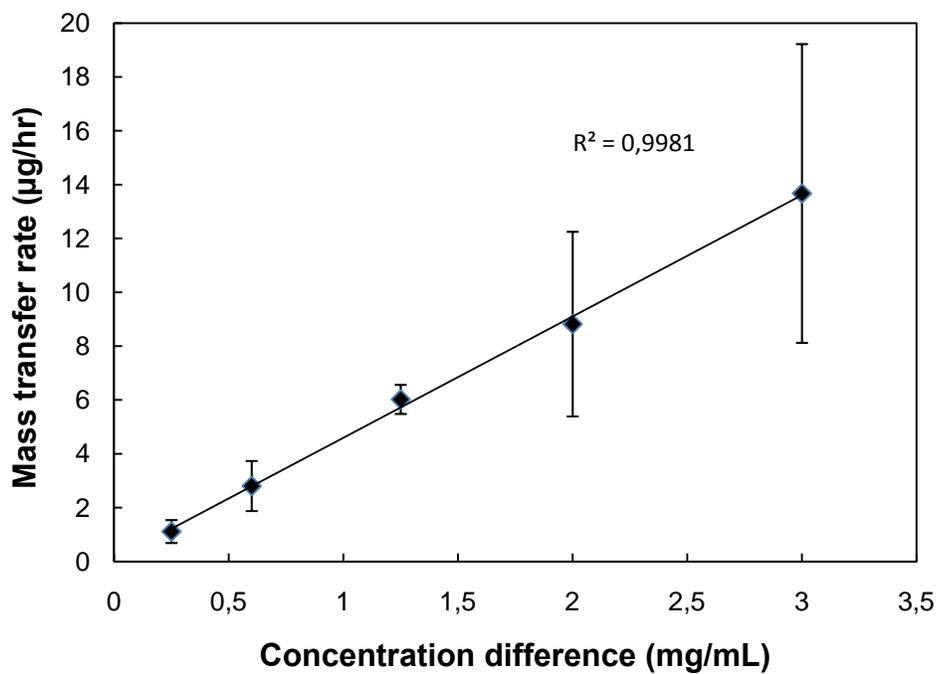

**Figure S1.** Sodium fluorescein mass transfer rate in the "on" state (45 °C), as a function of concentration difference across membrane. Data are for 25 wt% NG-37 membranes, with thermal triggering. Data are means ± SDs, n = 5 or 6 in each group.

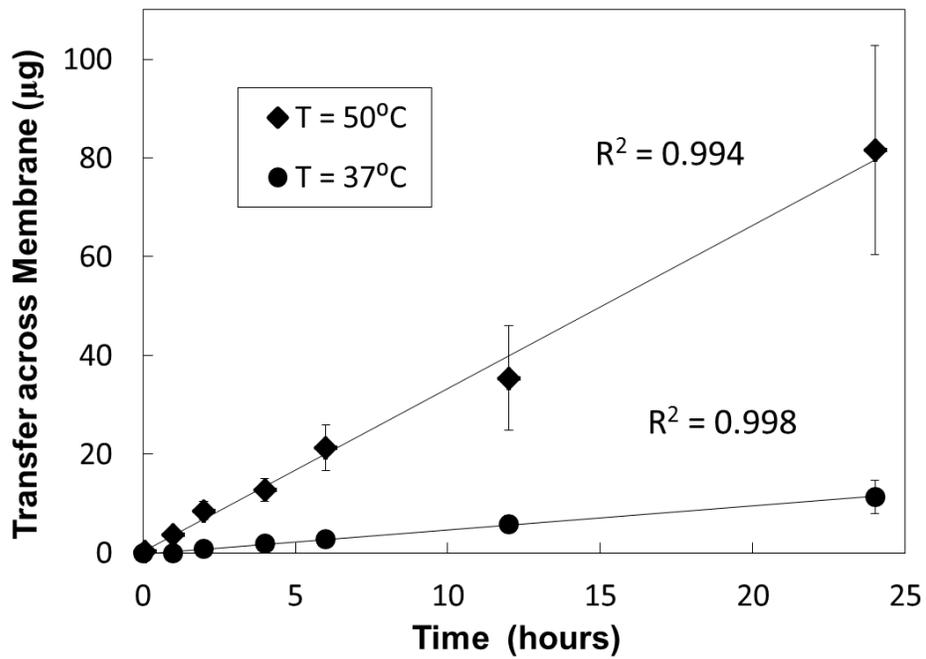

**Figure S2.** Release kinetics of sodium fluorescein from nanogel-filled magnetic membranes in the "off" (37° C) and "on" (50 °C) states. Data are for 25 wt% NG-42 membranes, with thermal triggering. Data are means ± SDs; n= 5 and n=4 for off and on states respectively.

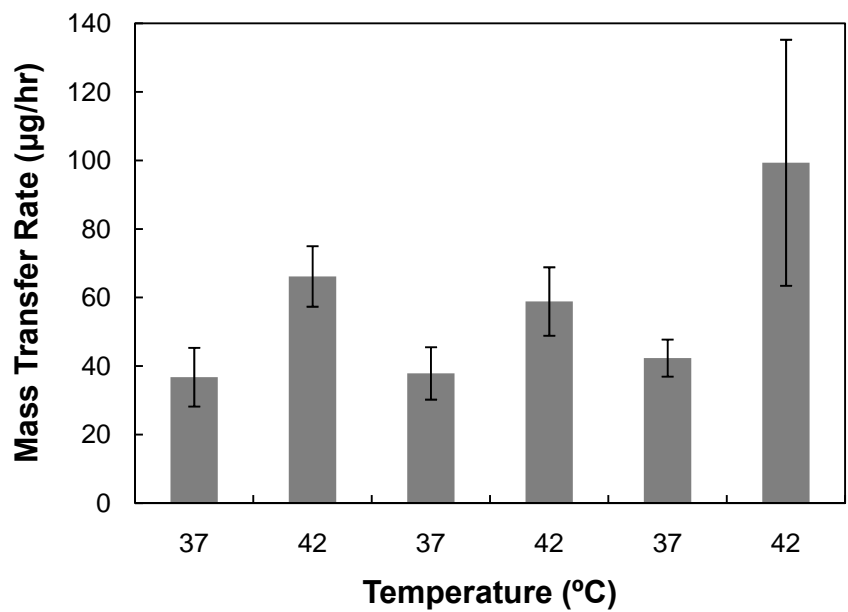

**Figure S3.** Release of bupivacaine from nanogel-filled magnetic membranes in the "off" (37 °C) and "on" (50 °C) states; 25 wt% NG-37 membrane, thermal triggering. Reservoir chamber was loaded with a saturated slurry of bupivacaine hydrochloride. Drug release was tracked by measuring the absorbance of the samples using a standard quartz cell and a Cary 50 UV/VIS spectrophotometer operating at 262nm. Data are means ± SDs; n= 5.

Effective temperature-dependent modulation of drug flux through the membrane was also observed for the cationic small molecule drug bupivacaine. However, the flux ratios between the on and off states were significantly lower than those achieved for anionic fluorescein, with a 1.9 ± 0.4-fold higher mass flux observed in the on state (42°C) relative to the off state (37°C). In comparison, for sodium fluorescein release through a similar membrane, a flux ratio of 8.0 ± 1.5 was observed (Figure 3). Ionic condensation between the nanogel-bound sulfate groups (derived from the persulfate initiator used to prepare the nanogels) and the cationic –$NH_2$ groups in bupivacaine may account for this lower on-off flux ratio. Charge-driven deswelling of the nanogel in the swollen state via ionic condensation by oppositely-charged small molecule drugs, previously reported with similar nanogels[12], could prevent the nanogel from fully swelling at low temperature and thus cause the membrane pores to remain slightly open even in the thermal off state, reducing the flux ratio as observed.

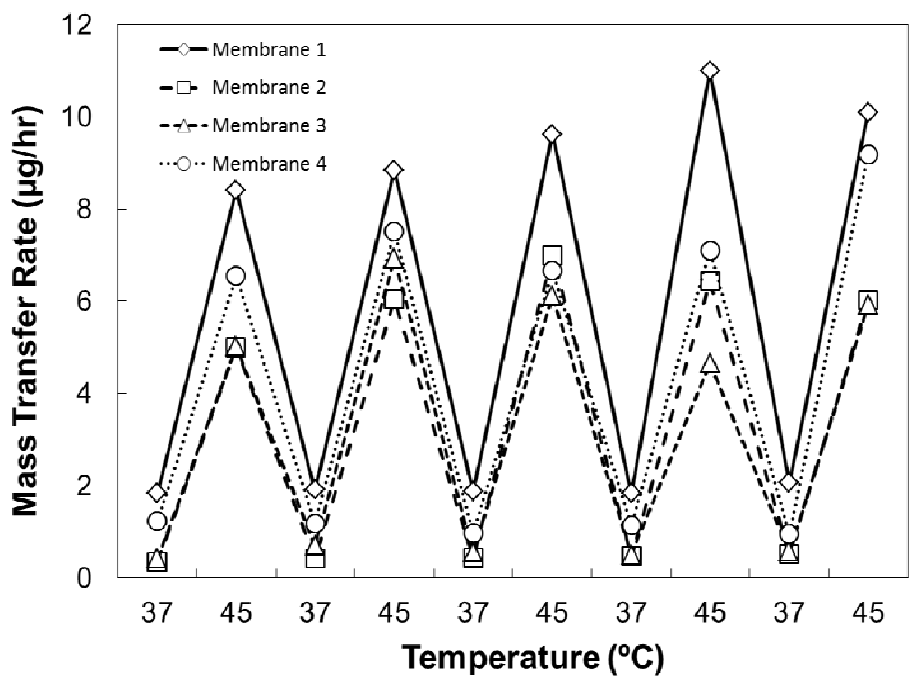

**Figure S4.** Membrane-to-membrane reproducibility for individual membranes represented in Figure 4. Data are for sodium fluorescein flux through thermally-triggered 25 wt% NG-37 membranes.

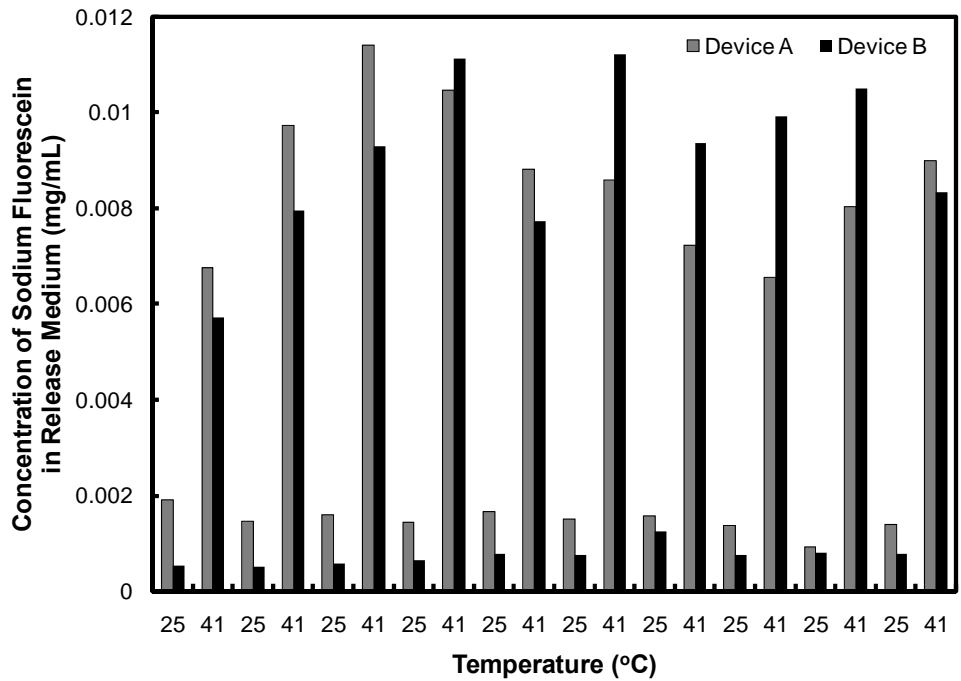

**Figure S5.** Reproducibility of drug release through membrane-capped devices. The capsules were loaded with 100 mg/mL sodium fluorescein and were triggered on and off for 10 cycles. 25 wt% NG-32 membrane; thermal stimulus. Membrane devices were fabricated as described in the Methods section for the magnetic triggering experiments.

The first cycle releases significantly less drug than subsequent cycles, likely attributable to the time required for the drug to fully saturate the gel phase inside the nanogel-filled pores of the membrane. Subsequent cycles release consistent doses of drug in each cycle; for example, for Device A, the average and standard deviation of the measured sodium fluorescein fluxes in the off and on states were 7 ± 1 μg/24 hours and 42 ± 6 μg/24 hours respectively over the subsequent nine cycles. In addition, the two devices tested exhibited no significant difference in release rates, with Device B releasing fluorescein at a rate of 47 ± 7 μg/24 hours in the on state ($p = 0.53$ comparing the on state fluxes of Device A and Device B). Thus, after the first triggering cycle, good reproducibility is observed both cycle-to-cycle and device-to-device.